# Tuning up or down the critical thickness in LaAlO$_3$/SrTiO$_3$ through *in situ* deposition of metal overlayers

By *D.C. Vaz*, *E. Lesne§*, *H. Naganuma*, *E. Jacquet*, *J. Santamaria, A. Barthélémy* and *M. Bibes**

[*]     Dr. M. Bibes, D.C. Vaz, Dr. E. Lesne, Dr. H. Naganuma, E. Jacquet, Prof. J. Santamaria, Prof. A. Barthélémy
Unité Mixte de Physique, CNRS, Thales, Univ. Paris-Sud, Université Paris-Saclay, 91767 Palaiseau, France

E-mail: manuel.bibes@cnrs-thales.fr

Dr. H. Naganuma

Tohoku University, Department of Applied Physics, 6-6-05 Aoba, Aramaki, Aoba, Sendai 980-8579, Japan

Prof. J. Santamaria
Universidad Complutense de Madrid

§ Present address: Max-Planck-Institut für Mikrostrukturphysik, Weinberg 2, 06120 Halle, Germany

The quasi two-dimension electron system (q2DES) that forms at the interface between LaAlO$_3$ (LAO) and SrTiO$_3$ (STO) has attracted much attention from the oxide electronics community[1]. Indeed, it has emerged as a platform to test a cornucopia of low-dimensional physical effects, such as two-dimensional superconductivity[2], quantum Hall effect[3], spin-charge conversion through the inverse Rashba-Edelstein effect[4] or electron pairing without superconductivity[5], to name just a few. The LAO/STO q2DES possesses several hallmark features; one is the existence of a critical LAO thickness of 4 unit-cells (uc) for interfacial conductivity to emerge[6]. Another is the extreme sensitivity of its transport properties to electrostatic boundary conditions. This surface-interface coupling was previously exploited to modulate both carrier densities and mobilities of the q2DES through the controlled adsorption of polar solvents[7–11] and by capping with different materials[12,13]. More spectacularly, first-principles calculations have suggested that the critical thickness could be reduced to just 1 uc by covering the LAO film with specific metals[14], an effect we recently demonstrated experimentally for Co[15]. However, the underlying mechanism leading to the formation of the q2DES in these Co/LAO(1 uc)/STO samples remains unclear.

In this paper, we investigate in detail the chemical, electronic and transport properties of several LAO(1-2 uc)/STO samples capped with different metals (Ti, Ta, Co, $Ni_{80}Fe_{20}$ – NiFe –, Nb, Pt, Pd and Au) grown in a ultra-high vacuum (UHV) system combining pulsed laser deposition (to grow the LAO), sputtering (to grow the metal) and *in situ* X-ray photoemission spectroscopy (XPS). The results confirm that for several metals a q2DES forms at 1-2 uc of LAO. Additionally, XPS shows that the appearance of interfacial conductivity is accompanied by a partial oxidation of the metal, a phenomenon that is strongly linked with the q2DES properties and with the formation of defects in this system. In contrast, for noble metals, the q2DES does not form at low LAO thicknesses and instead the critical thickness is increased *above* 4 unit cells. We discuss the results in terms of a hybrid mechanism that incorporates both electrostatic and chemical effects.

We first report on the magnetotransport response of LAO(2 uc)/STO structures capped with different metallic layers. We observe two very distinct types of Hall responses: for noble metals (Au, Pd, and Pt, *cf.* **Figure 1a**) the Hall trace is perfectly linear, and the Hall resistance only changes by a few tens of mΩ over a field of 9 T. The carrier densities associated with these tiny slopes are in the $10^{23}$ cm$^{-3}$ range, indicating that only the metal is contributing to transport in these samples. The Hall traces for LAO/STO structures capped with the other metals (Figure 1b) are very different: they are typically non-linear and the Hall resistance at 9 T is in the 100 Ω range. This indicates that besides the metal, an additional channel is contributing to transport i.e., a q2DES is formed in the STO. These results are qualitatively consistent with previous DFT calculations performed in metal/LAO/STO heterostructures[14], showing a vanishing Ti 3d band occupancy at the interfacial $TiO_2$ layer when capping an LAO(2 uc)/STO with one monolayer of Pt or Au, and a much larger one for Co and Ti.

In order to properly extract the transport response of the q2DES in these latter series of samples, we first analyze the longitudinal resistance measurements $R_{xx}$ as a function of magnetic field (magnetoresistance, MR). In this configuration, both conducting channels can be electrically described as two resistances in parallel ($R_M$ and $R_{q2DES}$), similarly to what was assumed in Ref. [15]. This relation is shown in Figure 1c for a LAO(2 uc)/STO sample capped with a thin Ta layer. Knowing the total contribution arising from both channels (black curve) and the single contribution of a Ta reference (grey curve) we are able to extract the signature of the q2DES (orange curve).

For the transverse resistance measurements, $R_{xy}$, besides having $R_M$ and $R_{q2DES}$ determining the current in each layer, the competition between the Hall voltages generated in both conducting channels must also be considered[4,16], as sketched in the inset of Figure 1b. We can then extract the q2DES Hall contribution (orange curve), represented in Figure 1d (where both Hall traces from the full stack (black curve) and the metallic layer (grey curve) are antisymmetrized). Here, the observation of an S-shaped Hall trace (accompanied by a strong MR) suggests multi-band type transport, already reported in previous publications[17–20]. To constrain the extracted parameters (carrier densities and mobilities), we fit simultaneously the Hall and the MR curves of the q2DES (light orange curve in Figure 1c and 1d) with a two-band model. For the Ta capped LAO(2 uc)/STO, we extract carrier densities of $n_1=3.59\times10^{13}$ cm$^{-2}$ and $n_2=0.49\times10^{13}$ cm$^{-2}$ with mobilities of $\mu_1=535$ cm$^2$/V.s and $\mu_2=2760$ cm$^2$/V.s.

We extract the Hall and MR traces of the q2DES formed in LAO(2 uc)/STO samples capped with NiFe, Co, Ti and Nb following the same procedure, cf. Figure 1e and 1f. For all cases, the Hall and the MR of the q2DES resemble those of the reference LAO(5 uc)/STO (dashed curve), namely an S-shape Hall trace and a positive MR. The precise shape of the Hall and MR curves varies with the metallic capping layer, suggesting different metal-oxide interactions.

To gain additional insight into the interaction between the LAO/STO and the metal, we used *in situ* X-ray photoemission spectroscopy (XPS). As reported in several previous studies[21–25], XPS can be used to probe the possible existence of a q2DES. Indeed, the presence of a conducting layer in the STO is associated with a valence change of the Ti 3d ions from 4+ (in the $3d^0$ bulk insulating state) towards 3+, detectable through a spectral intensity increase at the lower binding energy side of both spin-orbit split Ti 2p peaks.

In **Figure 2** we show the systematic XPS analysis of three LAO(1 uc)/STO samples capped with 3 Å of Co, Ta and Au. For comparison purposes, all spectra were normalized to the same integrated intensity and shifted in binding energy so that the Ti $2p_{3/2}$ peak maximum coincide in energy.

For the entire set, we started by looking at the Ti 2p core level spectra before and after the growth of a single crystalline LAO unit-cell. No additional $Ti^{3+}$ signal is observed after LAO deposition, confirming that no valence change has occurred (consistent with the absence of a q2DES for a LAO thickness smaller than the critical value of 4 uc [6] and with other spectroscopy studies[25,26]). We then transferred the sample in UHV to a sputtering chamber to deposit a thin layer of metal at room temperature, and transferred *in situ* the sample back to the XPS system. Depending on the metal overlayer, three different Ti 2p spectral changes were observed.

Figure 2a shows the results for a Co capped structure (green curve), for which a small $Ti^{3+}$ contribution arises. This observation is reminiscent of results on bare LAO(≥4 uc)/STO samples [8,21,23,24] and suggests that a q2DES is formed in this sample. To assess the depth distribution of the $Ti^{3+}$, we performed angle-dependent XPS experiments. While the 0° spectrum corresponds to the maximum probing depth, increasing the probing angle leads to a more interface sensitive measurement. We observe that at 50° the $Ti^{3+}$ signal is larger than at 0°, which indicates that the relative concentration of $Ti^{3+}$ is higher near the interface with the LAO (inset Figure 2a). This result is in line with what is reported in the literature for thick and uncapped LAO/STO[21], and matches with our transport experiments (Figure 1d) where the Hall and MR traces of both Co/LAO(2uc)/STO and LAO(5uc)/STO are very similar. Figure 2e shows a fit of the low binding energy part of the Ti $2p_{3/2}$ peak with $Ti^{3+}$ and $Ti^{4+}$ components, from which we estimate the $Ti^{3+}/Ti^{4+}$ peak area ratio to be ~1%.

For the Ta-capped heterostructure, cf. Figure 2b (red curve), a much stronger $Ti^{3+}$ contribution is visible, accompanied by an equally large decrease of the $Ti^{4+}$ signal. The magnitude of the $Ti^{3+}$ contribution clearly exceeds what has been observed for standard and uncapped LAO/STO above the critical thickness [8,23,24]. This suggests that the STO hosts an unusually large electron density after Ta deposition. As visible from the inset of Figure 2b, changing the probing angle does not lead to a significant $Ti^{3+}$ intensity difference, in contrast with the situation for Co. This indicates that the conducting region formed in these Ta/LAO/STO structures extends deeper than the XPS maximum probing depth (of about 5 nm[27]) and is thicker than in standard LAO/STO or Co-capped LAO/STO samples. Just as for the Co sample, we fit the Ti $2p_{3/2}$ peak with two components (cf. Figure 2g) which yields a $Ti^{3+}/Ti^{4+}$ peak area ratio of ~20%.

Lastly, the Au capped structure, represented in Figure 2c, shows no $Ti^{3+}$ contribution. Additionally, no additional $Ti^{3+}$ signal is evidenced through the XPS angle dependence (inset of Figure 2c). The only visible difference between the capped and uncapped sample is a slight broadening of the Ti $2p_{3/2}$ peak, which may be ascribed to potential structural damage induced by the deposition of heavy species such as Au. The data suggests that this sample does not host a q2DES, which is consistent with the Hall trace measured for this sample (Figure 1a). Here, the Ti $2p_{3/2}$ peak can be fitted with a single component corresponding to $Ti^{4+}$, so that the $Ti^{3+}/Ti^{4+}$ peak area ratio is ~0 (Figure 2i).

Figures 2d, 2f and 2h display the XPS spectra of each capping layer for the three samples. For Co, we compare the spectra collected on the sample (dark green curve) with references of pure metallic Co (black curve) and fully oxidized Co (light green curve). The Co/LAO/STO

data deviates from that of the metallic Co reference and shows signatures of oxidation. Using linear combinations of the metallic and oxidized Co reference spectra, we were able to estimate that 45±5% of the Co in the capping layer on top of the LAO/STO heterostructure was oxidized.

For the Ta capped sample (Figure 2d), the Ta spectrum corresponds to a mixture of Ta oxides[28], without indications of any metallic Ta, thus suggesting that nearly 100% of the Ta was oxidized. In contrast, the Au capping layer spectra appears identical to a metallic Au reference (Figure 2h), implying that no chemical reaction occurred with the LAO/STO structures.

Following Arras et al[14], it appears reasonable to assume at first that, for ideally stoichiometric heterostructures, the key parameter determining the formation of a q2DES in metal-capped LAO/STO samples is the work-function of the metal ($\phi_M$)[29]. Indeed, for metals with a low $\phi_M$, the Fermi level of the metal will initially be located higher in energy than the bottom of the conduction band of the STO, so that it is energetically favorable to transfer electrons from the metal to the STO, in turn leading to the q2DES formation. On the other hand, for metals with a high $\phi_M$, the Fermi level of the metal lies in the gap of the STO, so that no charge transfer occurs. Without the detailed insight brought by the *in situ* XPS experiments, this scenario appears to be valid since for low $\phi_M$ metals (such as Ta, Ti or Nb) a q2DES is formed while for high $\phi_M$ metals (Au, Pd and Pt) no interfacial conductivity is detected. However, XPS clearly displays signatures of metal oxidation, signaling the occurrence of a chemical reaction between the metal and the oxide structure. Thus, the physics and chemistry of the system cannot be fully described by a purely electrostatic model. In addition, defect formation and ionic diffusion must be taken into account[30].

In **Figure 3a-c** we plot the parameters extracted from the transport and XPS analysis as a function of the standard enthalpy of formation of metal oxide[29] $\Delta_f H^0$. It exhibits a trend where the amount of metal oxide deduced from Figure 2d, 2d and 2h scales with $\Delta_f H^0$ (Figure 3c).

Remarkably, this variation is correlated with the increase of the $Ti^{3+}/Ti^{4+}$ peak area ratio, as well as with the carrier density in the q2DES (see Figure 3a-b).

The scenario we propose to explain our experimental findings is summarized in Figures 3d (Co case), and 3e (Ta case). Prior to the metal deposition onto the LAO/STO, band tilting occurs due to the internal polar field in the LAO. When the metal is deposited, if the pure electrostatic criteria described previously is satisfied (i.e. the Fermi level of the metal lies higher in energy than the Fermi level in the STO) carriers are transferred from the metal towards the STO (so that the Fermi level on both sides align) and the q2DES is formed. Smaller $\phi_M$ translates in higher energy differences with respect to the Fermi level of the STO, so that larger charge transfer is promoted. This charge transfer creates a space charge field, $E_{SC}$, opposite to the polar field $E_P$ in the LAO, and which partially cancels the latter (depending on the amount of charge transferred).

Indeed, when comparing Co ($\phi_{Co}$=5 eV) with Ta ($\phi_{Ta}$=4.15 eV), the highest carrier density and the $Ti^{3+}/Ti^{4+}$ peak area ratio occurs for the Ta capped sample, consistent with the mechanism proposed. However, our XPS data show that while for Co the amount of transferred charge and the q2DES properties are similar to those of reference (uncapped) samples, for Ta, oxidation is much more severe. In turn the amount of transferred charge is much larger and, importantly, the q2DES has a larger spatial extension. This latter point is crucial, since in a purely electrostatic picture, a higher carrier density in the q2DES implies a stronger electric field in the STO and thus a stronger confinement[31], which is inconsistent with a thicker q2DES.

A way to reconcile these apparently contradictory observations is to assume that the extension of the q2DES is larger in the Ta-capped sample due to the presence of oxygen vacancies in the STO. This effect may in fact be reminiscent of the mechanism proposed by Yu and Zunger for uncapped LAO/STO[32] which states that at 4 uc of LAO oxygen vacancies spontaneously form at the LAO surface while the electrons associated with their formation are transferred to the STO and yield to a q2DES. Here, we argue that the deposition of a reactive metal *catalyzes* the formation of these oxygen vacancies. If the oxidation of the metal is moderate, as for the Co case, only a small amount of oxygen vacancies will be created in the LAO, where they will largely remain. The few electrons thus released will then be transferred to the STO and contribute carriers to the q2DES. However, if the oxidation of the metal is very severe as we observe for Ta ($\Delta_f H^0_{(Ta)}$= -2046 kJ/mol) the amount of oxygen vacancies created in the LAO will be much larger (as will the corresponding number of extra carriers transferred to the STO). For such high concentrations, these oxygen vacancies will tend to diffuse through the LAO layer and eventually reach the STO. There, because they are positively charged, oxygen vacancies will tend to be expelled from the q2DES region by the existing field and eventually generate an opposing electric field, $E_D$. This depolarization field weakens the confinement of the carriers in the STO near the interface, yielding an increase of the q2DES thickness (see Figure 3e).

Lastly, for the Au capped LAO/STO no hint of a q2DES was detected from transport, as well as no oxidation of the metallic overlayer from XPS. The weak electrochemical interactions between oxides and noble metals have previously been thoroughly investigated (see e.g. Ref. [30]), and mainly attributed to the high work function of the metals ($\phi_M$ >5 eV) and their reduced tendency to oxidize ($\Delta_f H^0 \approx 0$ kJ/mol).

To clarify how this interplay might affect the q2DES at LAO/STO interfaces we performed transport experiments on Pt(3 nm)/LAO(x)/STO heterostructures, where x=2, 5 and 10 uc (**Figure 4**). As in Figure 1a, the Pt/LAO(2uc)/STO sample shows a bulk metal-like Hall and MR behavior, indicating no q2DES formation. While for a LAO thickness of only 2 uc this result is expected *a priori*, no indication of a q2DES is detected for a Pt/LAO(5 uc)/STO sample either. In other words, the critical thickness is here *increased* beyond 4 uc.

To explain this puzzling result for the 5 uc sample, we postulate that prior to Pt deposition, a q2DES actually forms, assuming that electronic reconstruction already took place. Due to its high work function ($\phi_{Pt}$=5.64 eV), the Fermi level of Pt lies inside the gap of the STO.

After Pt deposition and band alignment (see Figure 4d), charge may then be transferred from the STO towards the Pt, depopulating the q2DES to the point where the interface becomes insulating. A similar effect has been reported for Pt/reduced-TiO$_2$, where the Ti$^{3+}$ signal of the substrate was substantially reduced after Pt deposition[33].

At a larger LAO thickness (10 uc), signatures of the q2DES eventually appear, as deduced from the change in the shape of the Hall, MR and normalized resistance versus *T* curves (see Figure 4a-c). Although the very low resistance of the Pt with respect to the q2DES in this case precludes an accurate determination of the carrier density, the data and the weak MR suggest that n$_s$ is here lower than for all the reactive metal-capped heterostructures discussed earlier[17]. The same line of thought proposed to explain the behavior of the 5 uc sample may be applied to the 10 uc sample. However, by increasing the LAO thickness, electron transfer towards the Pt may be hampered and therefore not sufficient to completely depopulate the q2DES, only leading to a reduction in the sheet carrier density. Alternatively, it is possible that as the LAO thickness increases from 5 to 10 uc, the electric field within the LAO also builds up which would eventually favor the transfer of electrons from the Pt to the STO.

In summary, we have shown that the electronic response of LAO/STO interfaces can be dramatically modified by the deposition of metallic overlayers. Not only can the carrier densities, mobilities and spatial extension be tuned by the metal, but the critical LAO thickness for q2DES formation – a hallmark feature of the LAO/STO – can also be decreased or increased from its nominal value of 4 unit cells, depending on the choice of capping material. This is illustrated in Figure 5. While electrostatics do play an important role to set the q2DES properties through band alignment effects, electrochemical reactions between the metal and the LAO also contribute by inducing oxygen vacancies in the LAO. If their density is high enough, their diffusion into the STO contributes electron carriers, as in reduced bulk STO crystals[34].

Our findings should stimulate further studies on the concomitant role of electrostatic boundary conditions and (electro)chemistry at metal/LAO/STO heterointerfaces, and on their influence on the emergence of the q2DES. A detailed understanding is necessary to enable controllable and reproducible properties at the device level, e.g. for efficient field-effect transistors in oxide-based electronics[35]. Beyond spin-injection from Co or NiFe into the q2DES through an ultrathin LAO film[4,36] new functionalities may arise from the combination of the q2DES with appropriate superconducting overlayers (e.g. electromagnetic coupling between two vortex fluids in Giaver-like transformer devices[37]), or with heavy metals with strong spin-orbit coupling[38].

**Experimental Section**

***Thin film growth.*** The LaAlO$_3$ (LAO) films were grown by PLD on 5 mm × 5 mm TiO$_2$-terminated (001)-oriented SrTiO$_3$ (STO) substrates (from Crystec GmbH). A single crystal LAO target was ablated by a KrF (248 nm) excimer laser at a repetition rate of 1 Hz and with a fluence of ~ 1 J/cm$^{-2}$. The LAO deposition was performed in an oxygen partial pressure of 2.0·10$^{-4}$ mbar and at a substrate temperature of 730°C. The substrate-to-target distance was set to 63 mm. After LAO growth, the samples were annealed for 30 min in about 400 mbar of oxygen at 500$^0$C. Finally, the LAO/STO heterostructures were cooled at 25$^0$C/min and kept in the same oxygen partial pressure for ~ 30 to 60 min. To prepare metal/LAO//STO samples, PLD growth was followed by *in vacuo* transfer and *in situ* deposition of a metal by dc magnetron sputtering at room temperature.

***X-ray photoemission spectrosopy*** was performed using a Mg K$_\alpha$ source (hv=1253.6 eV). Spectra analysis was carried out with the CasaXPS software.

***Transport measurements*** were performed with a Dynacool system from Quantum Design in the Van der Pauw configuration after bonding the samples with Al wires.

*Acknowledgements*

This work received support from the ERC Consolidator Grant #615759 "MINT", the region Île-de-France DIM "Oxymore" (project "NEIMO") and the ANR project "NOMILOPS". HN was partly supported by the EPSRC-JSPS Core-to-Core Program, JSPS Grant-in-Aid for Scientific Research (B) (#15H03548). D.C.V. thanks the French Ministry of Higher Education and Research and CNRS for financing of his PhD thesis. J.S. thanks the University Paris-Saclay (D'Alembert program) and CNRS for financing his stay at CNRS/Thales. The authors acknowledge fruitful discussions with H. Jaffrès, A. Sander, D. Preziosi, R. Claessen, P. Scheiderer and M. Sing.

Received: ((will be filled in by the editorial staff))
Revised: ((will be filled in by the editorial staff))
Published online: ((will be filled in by the editorial staff))

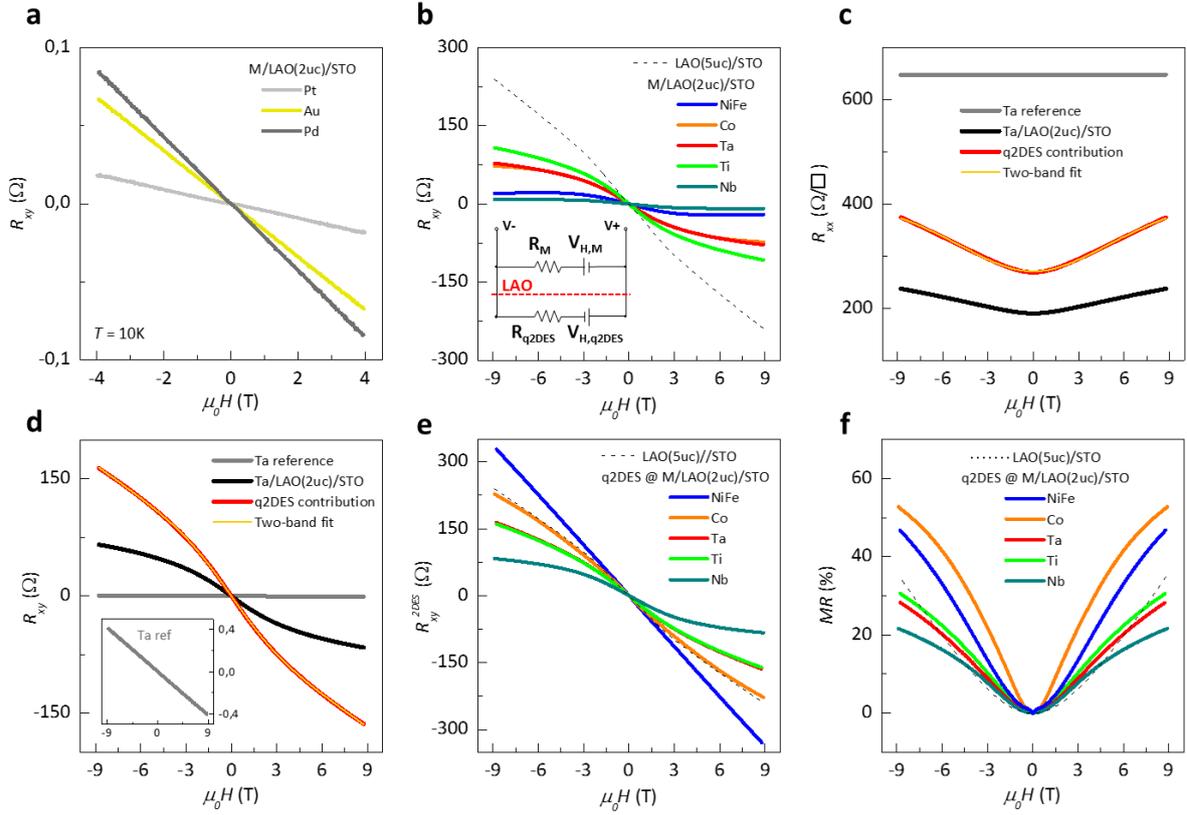

**Figure 1.** Antisymmetrized Hall resistance $R_{xy}$ as a function of applied perpendicular magnetic field $\mu_0 H$ for LAO(2 uc)/STO samples capped *in situ* with (a) noble metals (Pt, Au or Pd) and (b) with reactive metals ($Ni_{80}Fe_{20}$, Co, Ta, Ti or Nb). Data for an uncapped LAO(5 uc)/STO reference sample are added for comparison (dashed line). Inset: Schematic circuit after wire bonding of the whole structure, where $R_M$ and $R_{q2DES}$ represent the longitudinal resistances of the metal and q2DES conductive channels, respectively, and $V_{H,M}$ and $V_{H,q2DES}$ the Hall voltages generated in each channel while applying a magnetic field. (c) and (d) Extraction of the q2DES contribution (red curve) based on the magnetotransport data of both reference metal (grey curve) and metal+q2DES (black curve) for a Ta/LAO(2 uc)/STO sample. The obtained longitudinal resistance $R_{xx}$ and Hall data are fitted simultaneously with a two-band model (light orange curve). Inset: Zoomed Hall data for a reference Ta metallic layer. Hall resistance (e) and longitudinal magnetoresistance *MR* (f) of the q2DES formed at LAO(2 uc)/STO samples capped with reactive metals.

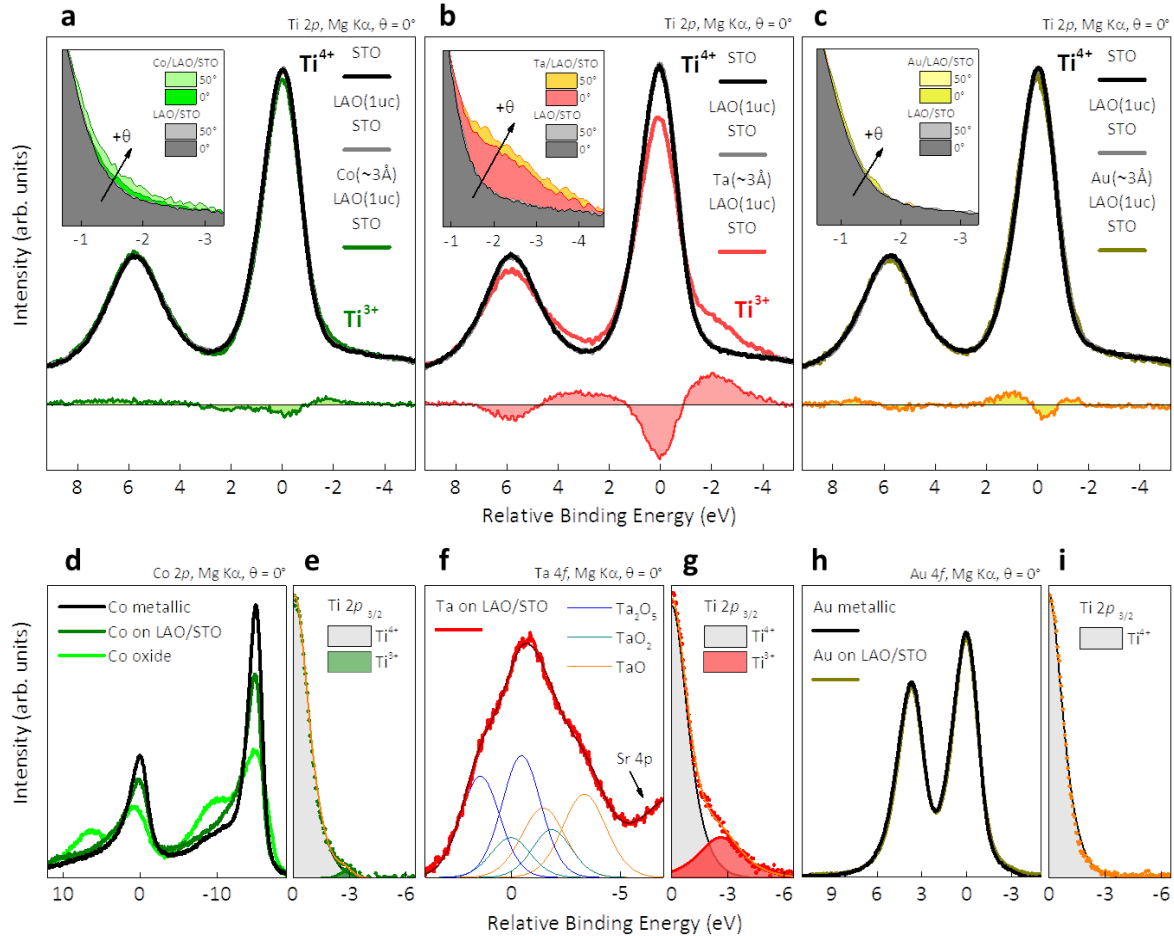

**Figure 2.** Normalized Ti 2p core-level spectra of LAO(1 uc)/STO samples capped by 3 Å of Co (a), Ta (b) and Au (c). For each sample, three stages are shown: STO bare substrate before deposition (black curve), after LAO growth (grey curve) and after metal deposition (colored curve). The difference between the grey and colored curves, shown below the spectra, highlights the presence of $Ti^{3+}$ after Co and Ta deposition. Insets: Angle dependence of the spectral weight in the $Ti^{3+}$ region before and after metal deposition. (e), (g) and (i) Expanded view of the Ti $2p_{3/2}$ core-level spectra (dots) fitted with both $Ti^{4+}$ and $Ti^{3+}$ contributions (shaded area). (d) Co 2p core-level spectrum of a Co(3 Å)/LAO(1 uc)/STO sample (green), compared with a fully metallic (black) and a fully oxidized (light green) Co reference sample, revealing a partially oxidized Co capping. (f) Fitted Ta 4f core-level spectrum of a Ta(3 Å)/LAO(1 uc)/STO sample with multiple Ta oxide peaks, following[28]. No Ta metal peak is discernable (usually

located 5 eV away from the oxide peak) which indicates that the Ta capping is nearly fully oxidized. (h) Au 4f core-level spectrum of a Au(3 Å)/LAO(1 uc)/STO. The overlap with the Au metallic reference hints to a fully metallic Au capping.

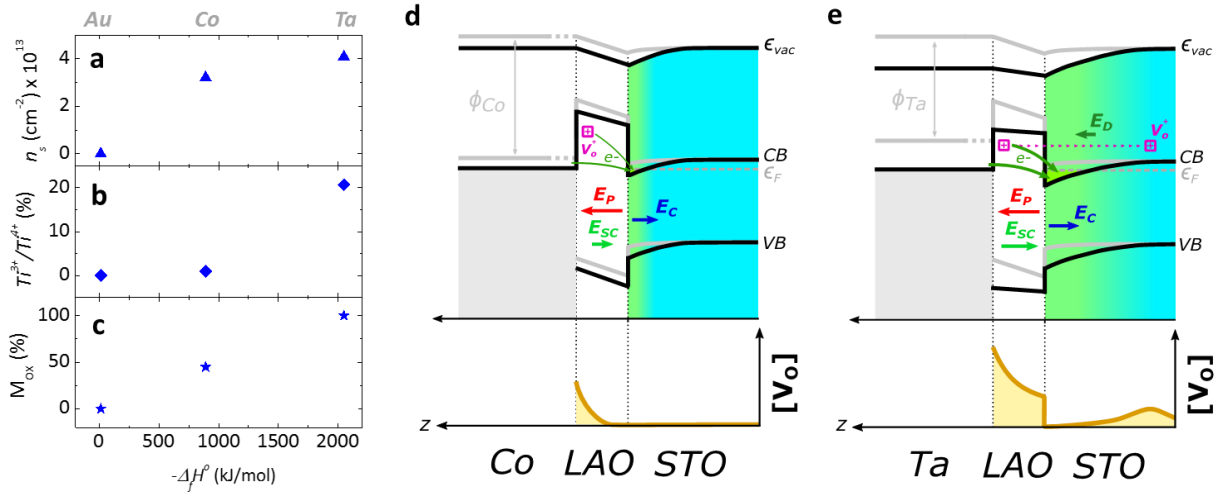

**Figure 3**. Sheet carrier density $n_s$ (a), $Ti^{3+}/Ti^{4+}$ peak area ratio (b) for the q2DES formed at M/LAO/STO heterostructures (M=Au, Co and Ta), and relative metal oxide quantity $M_{OX}$ (c) as a function of the enthalpy of oxide formation for Au, Co and Ta. Schematic energy band diagram for Co/LAO(1-2 uc)/STO (d) and Ta/LAO(1-2 uc)/STO (e) before (grey lines) and after (black lines) metal/LAO interface formation. Due to the difference in work functions, electrons are transferred from the metal to the STO, generating an electric field $E_{SC}$ opposing the polar field $E_P$ in LAO. Due to the oxidation of the metal at the interface with LAO, oxygen vacancies (pink squares) are formed in the LAO, providing more electrons also transferred to the STO. Electron transfer from the metal and due to oxygen vacancies formation to the STO are represented as dark green arrows. $E_C$ is the field in the q2DES region due to the band bending. The bottom part sketches the spatial profiles of the concentration of oxygen vacancies $[V_O]$ after diffusion. For Ta, the total amount of oxygen vacancies is much larger than for Co so that $[V_O]$ extends to the STO ; it vanishes in the q2DES region due to $E_C$. In the STO oxygen vacancies generate a field $E_D$ opposing $E_C$ (see text for details). CB and VB denote the conduction band minimum, and the valence band maximum respectively, $\epsilon_{vac}$ the vacuum level, and $\epsilon_F$ the Fermi energy.

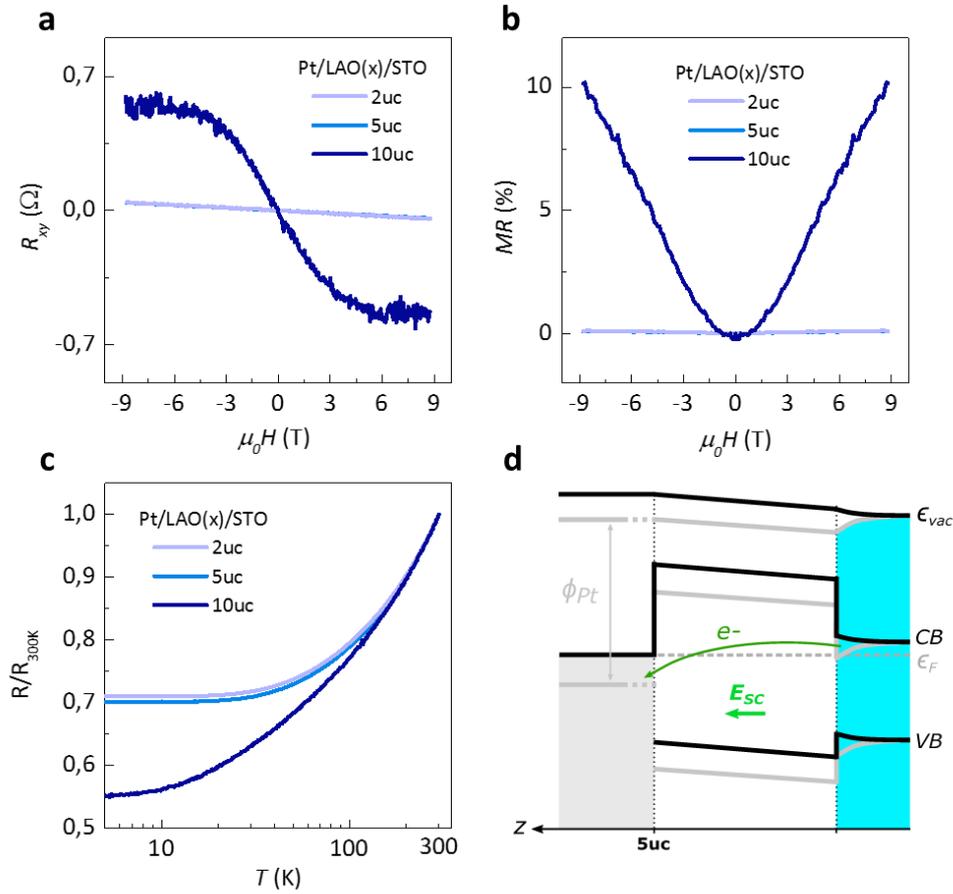

**Figure 4.** Hall resistance (a), MR (b), and normalized resistance as a function temperature *T* (c) for Pt(3 nm)/LAO(x)/STO samples, with x=2, 5 and 10 uc. (d) Schematic energy band diagram of Pt/LAO(5 uc)/STO before (grey lines) and after (black lines) Pt/LAO interface formation.

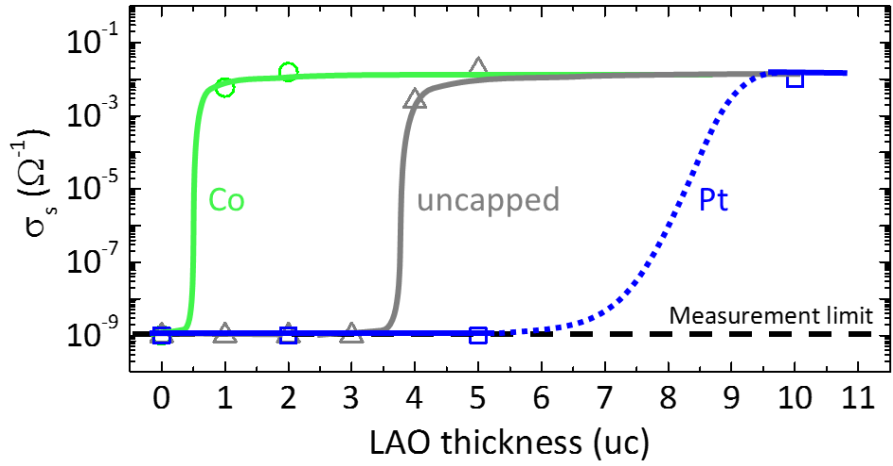

**Figure 5.** Sheet conductance of the q2DES *vs* LAO thickness in Co/LAO/STO, LAO/STO and Pt/LAO/STO samples at 10 K. The lines are guides to the eye.